\documentclass[preprint,review, 12pt]{elsarticle}
 
\usepackage{amssymb}

\usepackage{lineno}
\usepackage{graphicx}
\usepackage{subfigure}
\usepackage{multirow}
\usepackage{amsmath}
\usepackage{amssymb}
\usepackage{ulem}
\usepackage{verbatim}
\usepackage{upgreek}
\usepackage{xcolor}
\usepackage{makecell}
\usepackage{hyperref}

\usepackage{lineno}

\graphicspath{{fig/}}
\newif\ifcomment
\commenttrue
\commentfalse

\ifcomment
\usepackage[UTF8]{ctex}

\fi

\usepackage{xspace} 
\makeatletter
\ifcase \@ptsize \relax
  \newcommand{\miniscule}{\@setfontsize\miniscule{4}{5}}
\or
  \newcommand{\miniscule}{\@setfontsize\miniscule{5}{6}}
\or
  \newcommand{\miniscule}{\@setfontsize\miniscule{5}{6}}
\fi
\makeatother
\DeclareRobustCommand{\optbar}[1]{\shortstack{{\miniscule (\rule[.5ex]{0.75em}{.18mm})}
  \\ [-.7ex] $#1$}}
\def\porpbar    {\kern 0.18em \optbar{\kern -0.18em p}{}\xspace}

\journal{NIMA}

\begin{document}

\begin{frontmatter}

\title{\boldmath A conceptual design of TOF based on MRPC technology for the future electron-positron Higgs factory}

\author[a]{Kai Sun}
\author[b,c]{Yuexin Wang}
\author[a]{Jianing Liu}
\author[d]{Yongfeng Zhu}
\author[b]{Manqi Ruan\corref{cor1}}
\ead{ruanmq@ihep.ac.cn}
\author[a]{Yi Wang\corref{cor1}}
\ead{yiwang@mail.tsinghua.edu.cn}
\cortext[cor1]{Corresponding authors.}

\address[a]{Department of Engineering Physics, Tsinghua University, Beijing 10084, China}
\address[b]{Institute of High Energy Physics, Chinese Academy of Sciences, 19B Yuquan Road, Shijingshan District, Beijing 100049, China}
\address[c]{China Center of Advanced Science and Technology, Beijing 100190, China}
\address[d]{State Key Laboratory of Nuclear Physics and Technology, School of Physics, Peking University, Beijing, 100871, China}

\begin{abstract}

Future electron-positron Higgs factories could provide excellent opportunities to examine the Standard Model and search for new physics with much higher precision than the LHC.
A precise particle identification is crucial for the physics program at these future colliders and can be achieved via precise time-of-flight (TOF) measurements of the final state particles. 
In this paper, we propose a conceptual design of TOF system based on the multigap resistive plate chamber (MRPC) technology for future electron-positron Higgs factories.
This TOF system has a time resolution of $<$ 35\,ps, a total active area of 77\,m$^2$, and a construction budget of the order of 5 million USD.

\end{abstract}


\end{frontmatter}

\section{Introduction}
\label{sec:intro}
The discovery of the Higgs boson at the Large Hadron Collider (LHC)~\cite{ATLAS:2012yve,CMS:2012qbp} not only completes the particle spectrum in the Standard Model (SM), but also opens a new era of particle physics. Further measuring the properties of the Higgs boson with precision far beyond the LHC's becomes one of the primary goals of high energy physics. Compared with the hadron collider, the electron-positron collider, it is free of QCD background and can provide adjustable and measurable initial states, has the potential to improve the measurement precision of Higgs properties by at least one order of magnitude~\cite{CEPC_CDR_Phy,An:2018dwb}. Currently, there are four proposed electron-positron colliders worldwide that are intended to become future Higgs factories. These include the Circular Electron Positron Collider (CEPC)~\cite{CEPC_CDR_Acc,CEPC_CDR_Phy}, the Future Circular Collider $e^+e^{-}$ mode (FCC-ee)~\cite{FCC:2018byv,FCC:2018evy}, the International Linear Collider (ILC)~\cite{ILC_TDR_Sum,ILC_TDR_Phy,ILC_TDR_Acc,ILC_TDR_Det}, and the Compact LInear Collider (CLIC)~\cite{CLIC_CDR}. The CEPC is designed to have a circumference of 100\,km and two interaction points. Its center-of-mass energy spans a wide range from 91.2\,GeV to 360\,GeV. With the current high-luminosity design~\cite{Gao:2022lew}, as shown in table~\ref{tab:CEPCYield}, the CEPC will produce $3 \times 10^{12}$ $Z$ bosons, $1 \times 10^{8}$ $W^{+}W^{-}$ pairs, $4 \times 10^{6}$ Higgs bosons, and $5 \times 10^{5}$ $t\bar{t}$ pairs in total. The operation scheme of the FCC-ee~\cite{FCC:2018evy} is similar to that of the CEPC. Complementary with the circular collider, the linear collider that does not suffer from the synchrotron radiation can enable higher energy exploration. These future electron-positron colliders, which can measure the SM particles with unprecedented precision, provide unique opportunities to precisely examine the SM and to search for new physics.

\begin{table}[htbp]
    \centering
    \caption{The operation scheme of the CEPC. See \cite{Gao:2022lew} for details.}
    \label{tab:CEPCYield}
    \resizebox{.9\textwidth}{!}{
    \begin{tabular}[t]{ccccc}
        \hline
        Operation mode 
        & $Z$ factory & $W^+W^-$ & Higgs factory & $t\bar{t}$ \\
        \hline
        $\sqrt{s}$ (GeV) 
        & 91.2 & 160 & 240 & 360 \\
        Run time (year) 
        & 2 & 1 & 10 & 5 \\
        \makecell{Instantaneous luminosity \\ ($10^{34}\,{\rm cm}^{-2}{\rm s}^{-1}$, per IP)} 
        & 191.7 & 26.6 & 8.3 & 0.83 \\
        \makecell{Integrated luminosity \\ (${\rm ab}^{-1}$, 2~IPs)} 
        & 100 & 6 & 20 & 1 \\
        Event yields 
        & $3 \times 10^{12}$ & $1 \times 10^{8}$ & $4 \times 10^{6}$ & $5 \times 10^{5}$ \\
        \hline
    \end{tabular}
    }
\end{table}

The precise time-of-flight (TOF) measurement is essential to achieve the scientific goals of these future collider projects. The most common application of TOF in high energy physics experiments is the particle identification (PID), which generally refers to the discrimination of $\pi^{\pm}/K^{\pm}/p$. An efficient PID is particularly vital for separating decays with similar topologies in final states, such as $B^0_{(s)}\to\pi^+\pi^-$, $B^0_{(s)}\to K^+K^-$, and $B^0_{(s)}\to K^{\pm}\pi^{\pm}$. These decays play important roles in measurements of the Cabibbo-Kobayashi-Maskawa (CKM) matrix~\cite{LHCb:2012ihl}. In addition, the excellent PID performance is also critical for the jet charge measurement~\cite{Hanhua_report_2022} that is necessary for $CP$ violation measurements. These are all basic and important parts of flavor physics that can be thoroughly investigated at the $Z$-pole operation of CEPC and FCC-ee. For wide momentum ranges of these future colliders, TOF information can well complement the $dE/dx$ measurement and significantly improve the PID performance in the low momentum range ($<$ 4\,GeV)~\cite{CEPC_CDR_Phy,An:2018jtk,Zhu:2022hyy}. The required TOF resolution is at least 50\,ps~\cite{An:2018jtk,Zhu:2022hyy}. Besides the PID, the TOF measurement also shows considerable potential in resolving some ambiguities, such as jet confusion, off-time pileup, and confusion in the particle flow reconstruction (e.g. fragmented clusters from charged hadrons misidentified as neutral hadrons~\cite{ruan2019performance}).

\begin{table}[htbp]
    \centering
    \caption{Key parameters of MRPC adopted to construct TOF systems in high energy physics experiments.}
    \label{tab:MRPCinHEPToFSystem}
    \resizebox{1.\textwidth}{!}{
    \begin{tabular}[t]{cccccc}
        \hline
        Experiments
        & \makecell{Total active area\\ (m$^2$)}  & \makecell{Gap$\times$thickness\\ (mm)} & \makecell{Max rate\\ (Hz/cm$^2$)} & \makecell{Efficiency\\ (\%)} & \makecell{Time resolution\\ (ps)} \\
        \hline
        STAR~\cite{wang2010production}
        & 50 & 6$\times$0.22 & 10 & 95--97 & 60 \\
        ALICE~\cite{akindinov2006mrpc}
        & 141 & 10$\times$0.25 & 50 & 99.9 & 40 \\
        BES\uppercase\expandafter{\romannumeral3}~\cite{wang2016upgrade}
        & 1.33 & 12$\times$0.22 & 50 & 99 & 60 \\
        CBM~\cite{wang2016development}
        & 120 & 10$\times$0.25 & 30k & 97 & 60 \\
        SoLID ~\cite{yu2022development}
        & 10 & 32$\times$0.128 & 20k & 98 & 20 \\
        \hline
    \end{tabular}
    }
\end{table}

Among all existing TOF technologies, the multigap resistive plate chamber (MRPC) features high detection efficiency, fast timing response, and high rate capability. It has been widely used as the TOF system in many high energy physics experiments, such as BESIII~\cite{wang2016upgrade}, ALICE~\cite{akindinov2006mrpc}, STAR~\cite{wang2010production}, CBM~\cite{wang2016development}, and SoILD~\cite{yu2022development}. Table~\ref{tab:MRPCinHEPToFSystem} summarizes the key parameters of MRPC-based TOF systems (abbreviated as MRPC-TOF hereafter) in these experiments. The MRPC-TOF in the CBM experiment~\cite{wang2016development} is an excellent instance of high counting rate MRPC that can operate at a counting rate up to 30\,kHz/cm$^2$ with an efficiency of 97\% and a time resolution of 60\,ps. In addition, an impressive time resolution better than 20\,ps can be achieved using MRPCs with gas gaps as narrow as 128\,$\upmu$m in the SoILD experiment~\cite{yu2022development}. The above excellent performance of MRPC shows its considerable potential to be applied in future electron-positron colliders and to fulfil their stringent requirements for precise time measurements and high radiation environments. On the other hand, from the perspective of construction, the MRPC is robust, inexpensive, and easy to fabricate compared to other TOF technologies, especially for large detector systems. Considering these advantages of MRPC, we therefore propose a conceptual design of MRPC-TOF system for future electron-positron colliders in this paper.

The remainder of this paper is organized as follows.
A detailed introduction of the MRPC technology, especially our independently developed sealed MRPC design with a time resolution of $<$ 20\,ps, is presented in section~\ref{sec:MRPC}. The conceptual design of the whole MRPC-TOF system is described in section~\ref{sec:Design}, using the CEPC as an example. A conclusion is drawn in section~\ref{sec:Conclusion}.

\section{MRPC technology}
\label{sec:MRPC}

The MRPC is a type of high-performance gas detector that is both cost-effective and easy to manufacture. In addition to the advantages of the MRPC technology described in section~\ref{sec:intro}, this section will focus on our independently developed MRPC with a time resolution of $<$ 20\,ps and a specially designed sealing structure.

The basic structure of the MRPC is relatively simple. It consists of several resistive plates separated by fishing lines, forming multiple gas gaps. The resistive plates are equipped with high-voltage electrodes on their outer surfaces, which create an electric field of over 100\,kV/cm within the gas gaps. When high-energy charged particles pass through the gas gaps, the working gas within the gaps is ionized, causing an avalanche effect that generates many electrons. These secondary electrons then drift through the gas gaps, inducing a high-speed signal on the readout strips.

\begin{figure}[htbp]
    \centering
    \includegraphics[width=.7\textwidth]{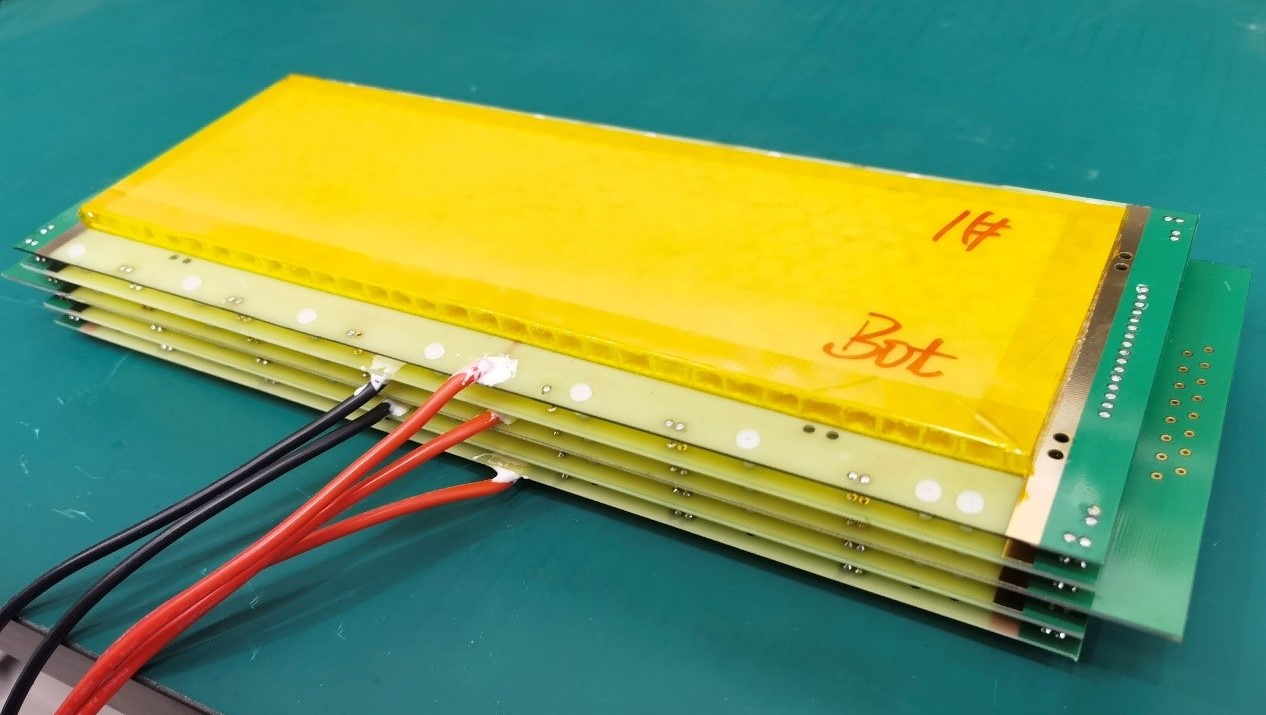}
    \caption{\label{20psMRPCStructure2}Structure of high time resolution MRPC.}
\end{figure}

\begin{figure}[htbp]
    \centering
    \subfigure[Time resolution before event selection.]{
    \label{Fig.sub.1}
    \includegraphics[width=.7\textwidth]{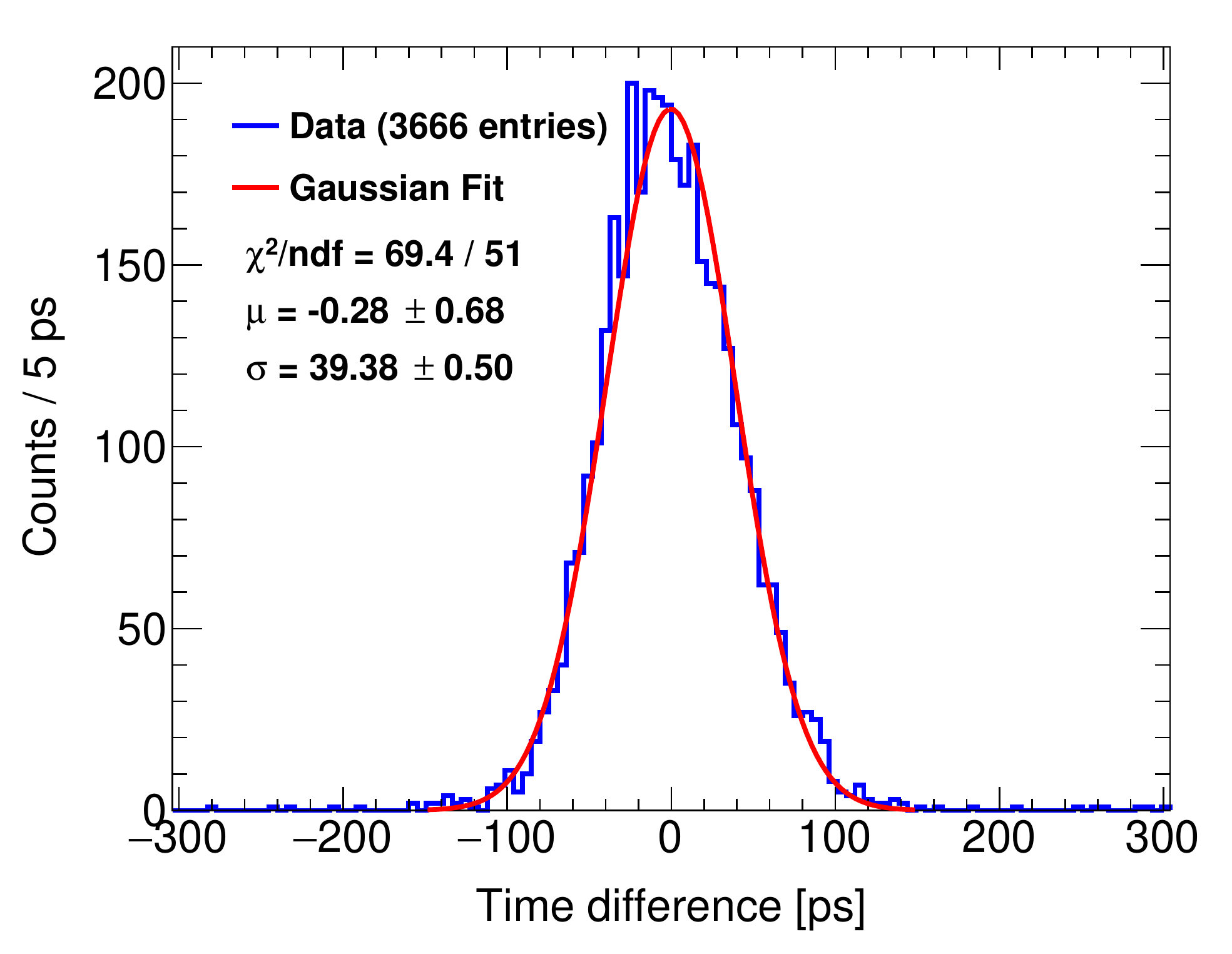}}
    \subfigure[Time resolution after event selection.]{
    \label{Fig.sub.2}
    \includegraphics[width=.7\textwidth]{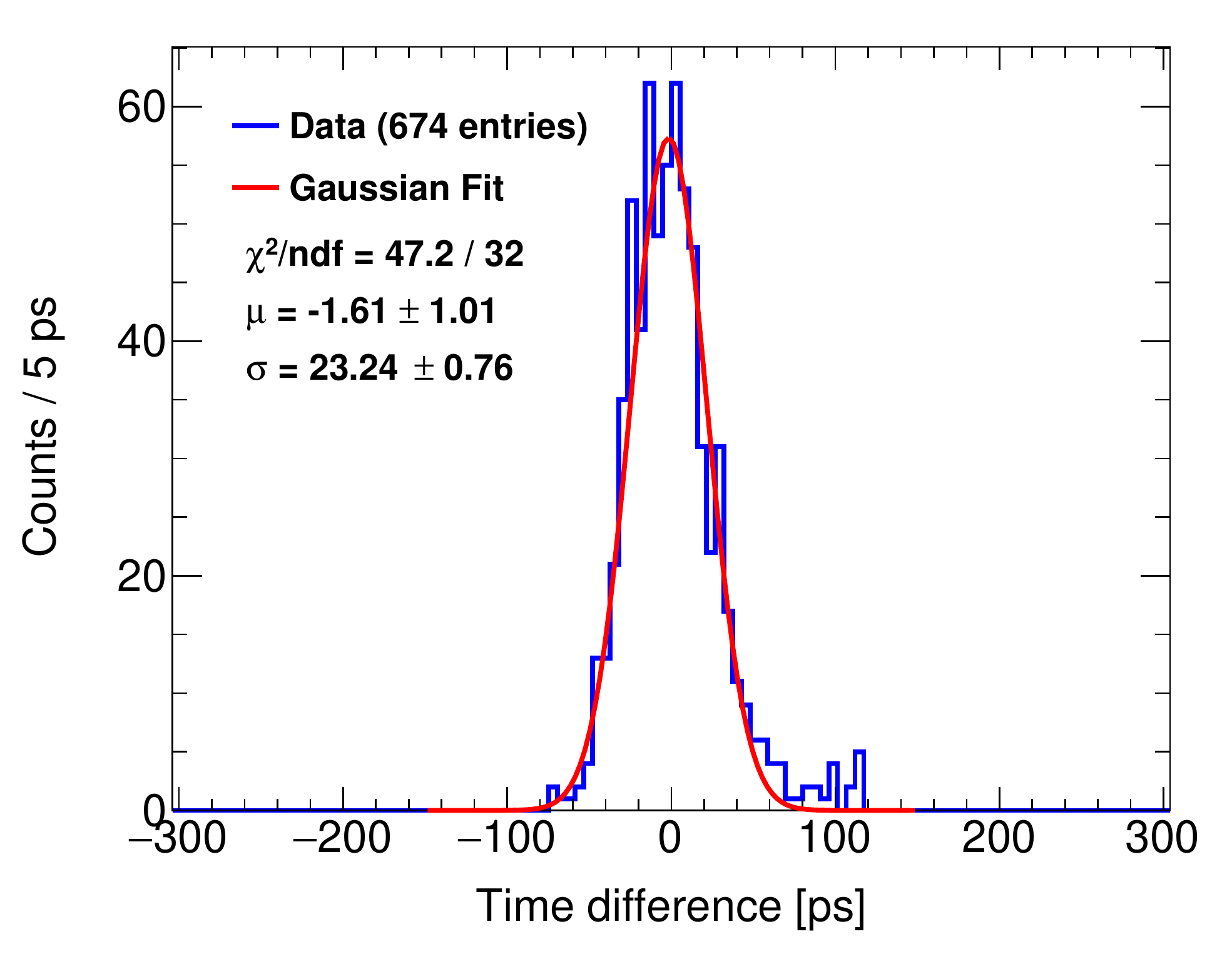}}
    \caption{\label{TimeResolution20ps}Time resolution of cosmic ray test.}
\end{figure}

In our recent research, we have developed an MRPC that can achieve a time resolution of 17\,ps~\cite{yu2022development}. Figure~\ref{20psMRPCStructure2} shows the detector prototype: there are 4 chambers and 32 gas gaps in an MRPC, and the gas gap width is 128\,$\upmu$m. The high voltage is supplied by a carbon film that is sprayed onto the surface of the outer glass.
For the cosmic ray test system, two vertically arranged scintillator detectors are used as triggers to confirm that the approximately vertical cosmic ray passes through the two MRPC detectors. When the avalanche effect happens, the signals on readout strips are amplified by the Analog Frontend Electronics (AFE)~\cite{liu2019design} from the University of Science and Technology of China (USTC), which have a time jitter smaller than 5\,ps. The times the cosmic ray flies through two MRPCs are then recorded by the waveform digitization module based on the DRS4 chips, to calculate the time resolution. The wave sampling board can reach a time resolution of 10\,ps, and it has a sampling frequency of 5.12\,GHz.
The final time resolution of the single MRPC detector is given by formula~\eqref{SigmaMRPC}, where $t_{\rm TL}$ and $t_{\rm TR}$ are the timings of the left and right strips for the top MRPC, and $t_{\rm BL}$ and $t_{\rm BR}$ are the same quantities for the bottom MRPC.
\begin{equation}
    \sigma_{\rm MRPC}=\frac{\Delta t}{\sqrt{2}},\ \Delta t=\frac{(t_{\rm TL}+t_{\rm TR})}{2} - \frac{(t_{\rm BL} +t_{\rm BR})}{2}
    \label{SigmaMRPC}
\end{equation}
Figure~\ref{TimeResolution20ps} presents the results of the cosmic ray test. At a high voltage of 13.3\,kV, a time resolution of $39.38\,{\rm ps}/\sqrt{2}\approx28\,{\rm ps}$ is obtained.  After selecting the events with approximately vertical cosmic rays, a time resolution of up to $23.24\,{\rm ps}/\sqrt{2}\approx17\,{\rm ps}$ can be achieved, which can fully fulfill the requirements of future Higgs factories for the 50\,ps TOF resolution.

\begin{figure}[htbp]
    \centering
    \subfigure[Structure of sealed MRPC.]{
    \label{Fig.sub.1}
    \includegraphics[width=.9\textwidth]{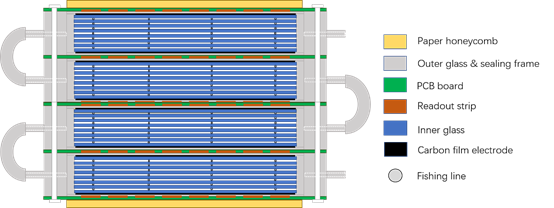}}
    \subfigure[Sealed MRPC in kind.]{
    \label{Fig.sub.2}
    \includegraphics[width=.7\textwidth]{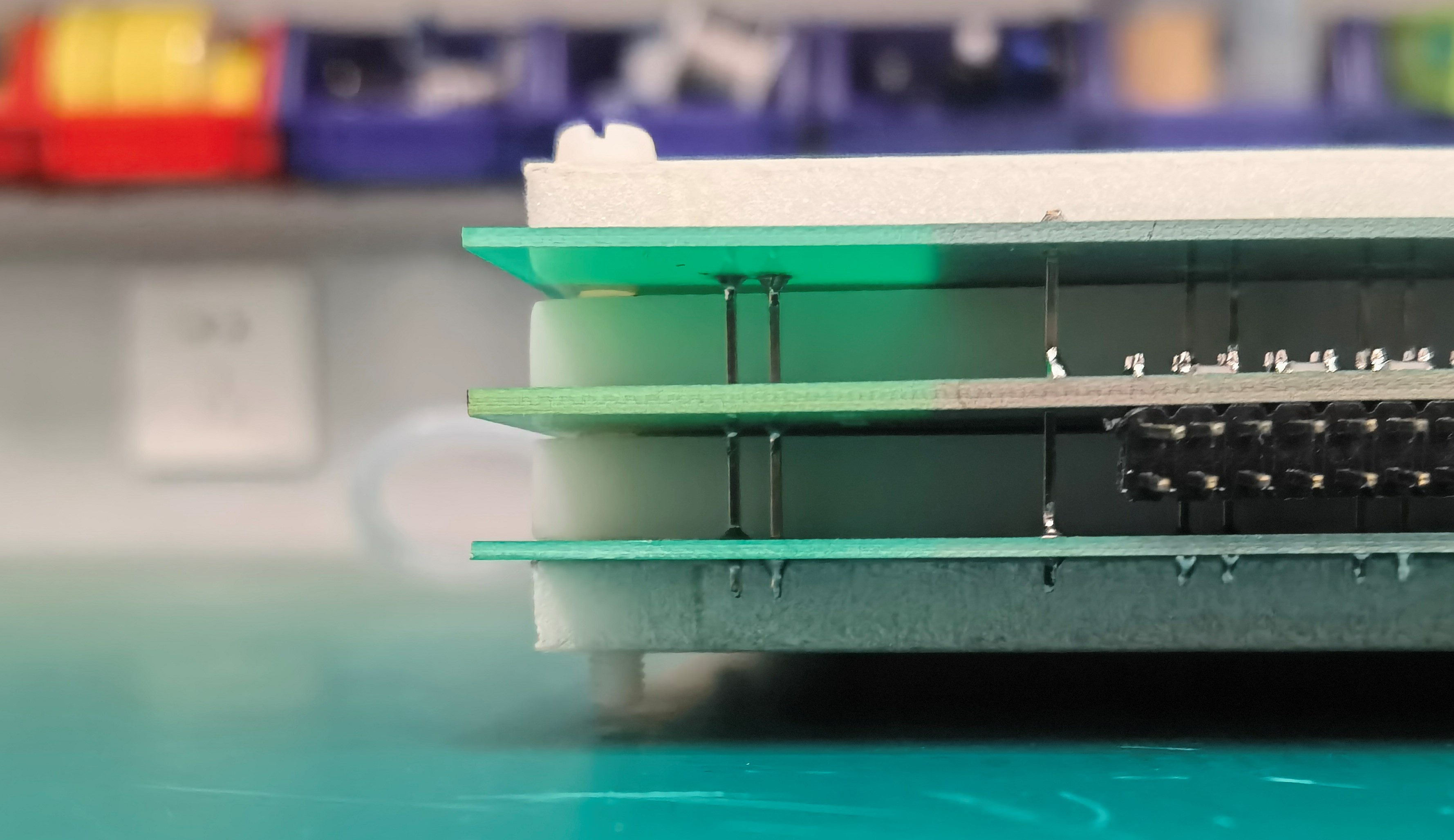}}
    \caption{\label{SealedMRPC3}Sealed MRPC.}
\end{figure}

As mentioned above, MRPC is a gas detector based on the principle of avalanche amplification. To prevent excessive avalanche spread, electronegative gases C$_2$H$_2$F$_4$ (Tetrafluoroethane), SF$_6$ (sulfur hexafluoride), and photon-absorbing gas i-C$_4$H$_{10}$ (iso-butane) are typically used in MRPC. However, Tetrafluoroethane and sulfur hexafluoride are considered as greenhouse gases, and efforts are underway to reduce their use. Over the past two years, our laboratory has developed a sealed MRPC design~\cite{chen2020development,wang2021high} that can significantly reduce greenhouse gas consumption. In this design, each chamber is enclosed within an airtight structure containing glass, gas gaps, and fishing lines. These components, which are typically exposed in traditional MRPCs, are surrounded by a completely sealed frame~\cite{wang2020cee}, leaving only the inlet and outlet gas ports on both sides, as shown in figure~\ref{SealedMRPC3}. This configuration can significantly reduce the total gas volume and increase gas exchange efficiency. During experiments, stable operation with a gas flow rate of 5 sccm/m$^2$ can be achieved. The gas flow rate per unit area of the sealed MRPC is only 1/10 of that of traditional MRPCs, but the detector's performance is maintained. The global warming potential (GWP) is used to evaluate the greenhouse effect of gases. It is defined as the heat absorbed by a given greenhouse gas in the atmosphere, as a multiple of the heat that would be absorbed by the same mass of CO$_2$. The GWPs of MRPC gases are shown in table~\ref{tab:GWPs}. Using traditional MRPC, the CEPC-TOF system covering an area of 77\,m$^2$ consumes 2023.6\,m$^3$ of gas mixture per year, i.e. 7740.2\,kg of C$_2$H$_2$F$_4$ and 624.3\,kg of SF$_6$. The sealed MRPC can reduce 90$\%$ emission of greenhouse gases, and therefore, reduce the GWP down to 2.339$\times$10$^4$ tonnes of CO$_2$ each year. The use of sealed MRPCs in the TOF system of future Higgs factories can help reduce the greenhouse gas consumption and promote overall environmental protection in the experimental device.

\begin{table}[htbp]
    \centering
    \caption{GWPs and densities of gases in MRPC }
    \label{tab:GWPs}
    \resizebox{.6\textwidth}{!}{
    \begin{tabular}[t]{ccc}
        \hline
        Gas type 
        & GWP (100 years)  & Density (kg/m$^3$)\\
        \hline
         CO$_2$ 
        & 1 & 2.00\\
         C$_2$H$_2$F$_4$
        & 1430 & 4.25\\
        SF$_6$
        & 23900 & 6.17\\
        \hline
    \end{tabular}
    }
\end{table}

\section{Conceptual design of MRPC-TOF for CEPC}
\label{sec:Design}

\begin{figure}[htbp]
\centering
\includegraphics[width=.75\textwidth]{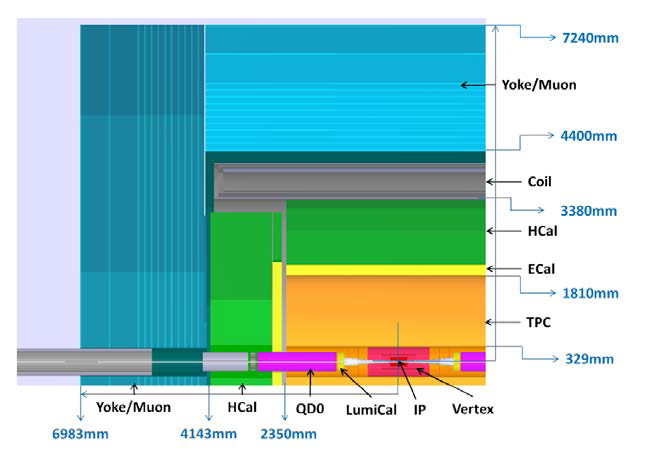}
\caption{\label{det}The layout of the CEPC baseline detector design.}
\end{figure}

Based on the above modular and environmentally friendly sealed MRPC design, the conceptual MRPC-TOF design is presented in this section. It is proposed for future electron-positron colliders, taking the CEPC as an example. The layout of the CEPC baseline detector~\cite{CEPC_CDR_Phy} is shown in figure~\ref{det}. The entire detector design follows the particle flow principle~\cite{Ruan:2013rkk}, emphasizing the efficient separation of final state particles and precise measurement of each final state particle in the most appropriate subdetector. From innermost to outermost, the baseline detector concept consists of a silicon pixel vertex detector, a silicon inner tracker, a TPC surrounded by a silicon external tracker, a silicon-tungsten sampling electromagnetic calorimeter (ECAL), a steel-glass resistive plate chambers sampling hadronic calorimeter (HCAL), a 3-Tesla superconducting solenoid, and a flux return yoke embedded with a muon detector. The proposed MRPC-TOF is designed to be installed in the inner layer of the ECAL.

\begin{figure}[htbp]
    \centering
    \includegraphics[width=.7\textwidth]{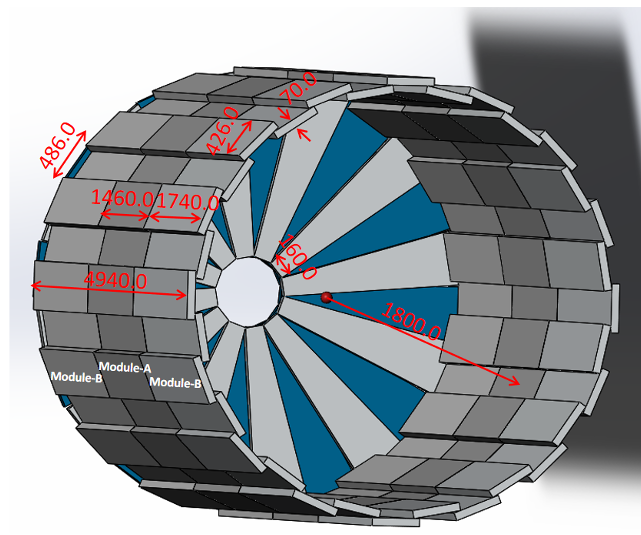}
    \caption{\label{MRPC_TOF}The configuration of the TOF system.}
\end{figure}

The detailed configuration of the MRPC-TOF system is shown in figure~\ref{MRPC_TOF}. It consists of one cylindrical barrel and two plate endcaps. The inner radius and length of the barrel are 1800\,mm and 4940\,mm, respectively. The total area of the whole TOF system is about 77\,m$^2$.

\begin{figure}[htbp]
    \centering
    \includegraphics[width=1.\textwidth]{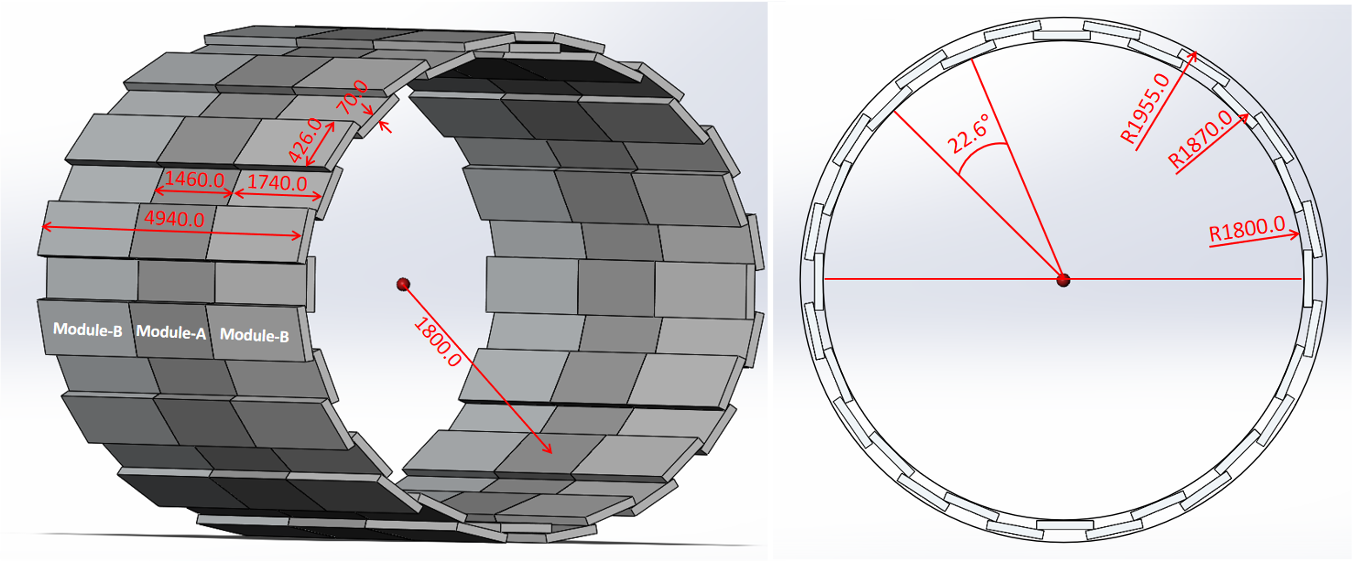}
    \caption{\label{MRPC_Barrel}The configuration of the TOF barrel.}
\end{figure}

\begin{figure}[htbp]
    \centering
    \includegraphics[width=1.\textwidth]{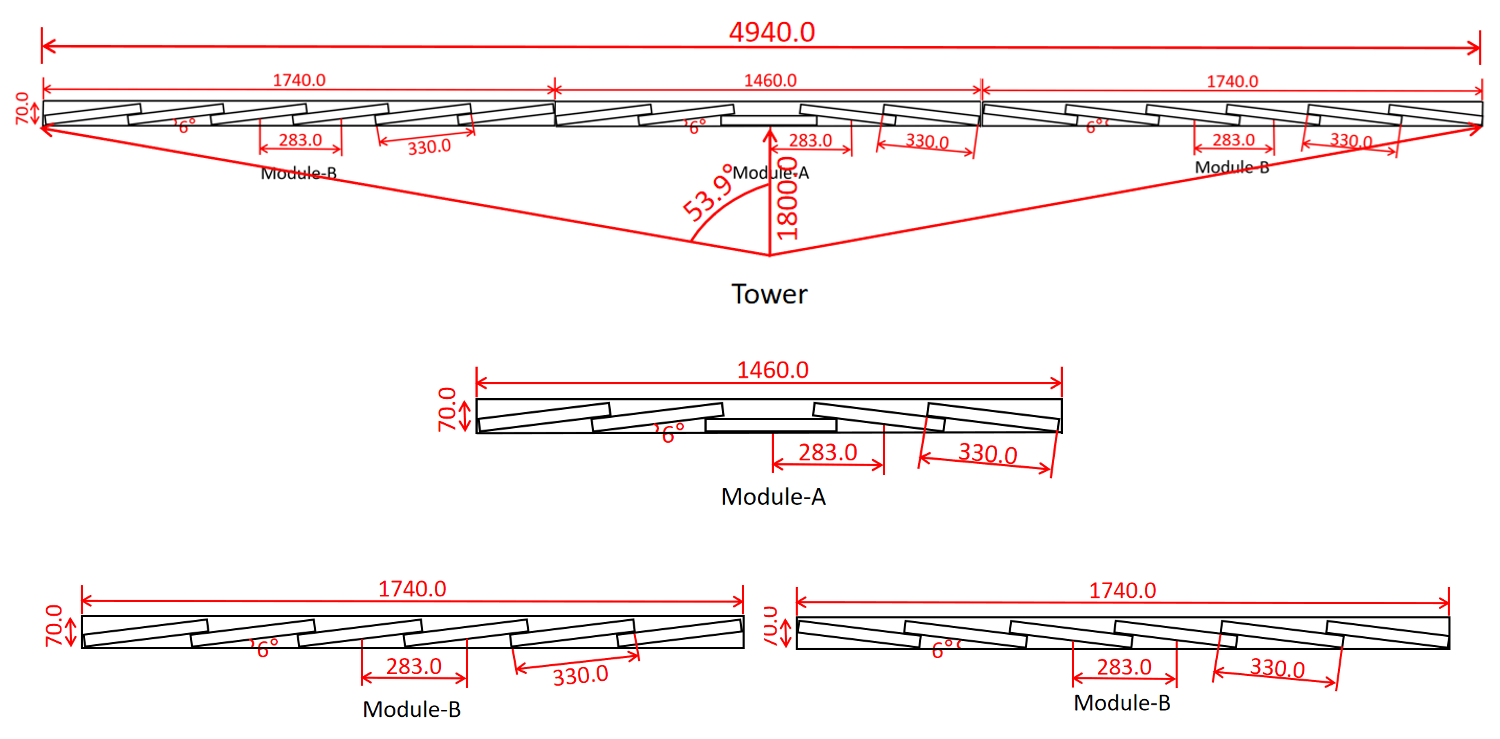}
    \caption{\label{MRPC_Tower}Arrangement of MRPCs inside the box along the beam direction.}
\end{figure}

For the barrel region, as shown in figure~\ref{MRPC_Barrel}, two layers of towers overlap along the circumferential direction, with 16 towers per layer. Each tower contains one Module-A and two Module-B with detailed structures shown in Figure~\ref{MRPC_Tower}. There are two overlapping readout strips between two neighboring MRPCs in Module-A and Module-B, with a width of 10+2.5\,mm. 

\begin{figure}[htbp]
    \centering
    \includegraphics[width=1.\textwidth]{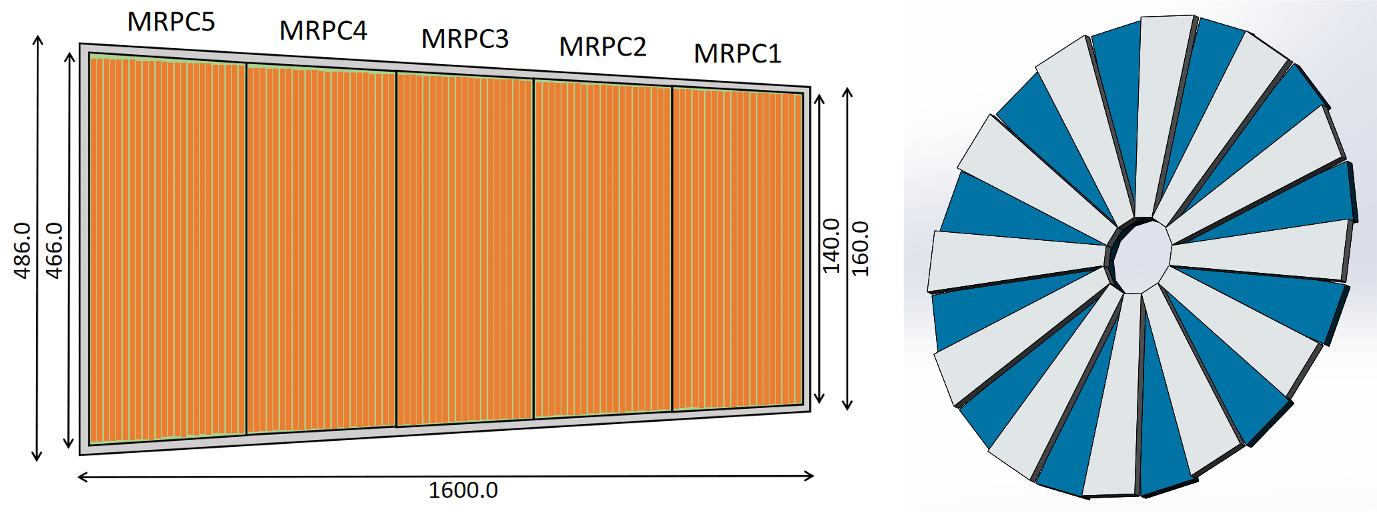}
    \caption{\label{MRPC_Endcap}The configuration of the TOF endcap.}
\end{figure}

As shown in figure~\ref{MRPC_Endcap}, there are a total of 24 modules at one endcap, which overlap in two layers. Each module has five MRPC detectors, with 1-2 overlapping readout strips between two neighboring MRPCs. Each detector has 24 readout strips, which are 10+3.5\,mm in width.

\begin{figure}[htbp]
    \centering
    \includegraphics[width=1.\textwidth]{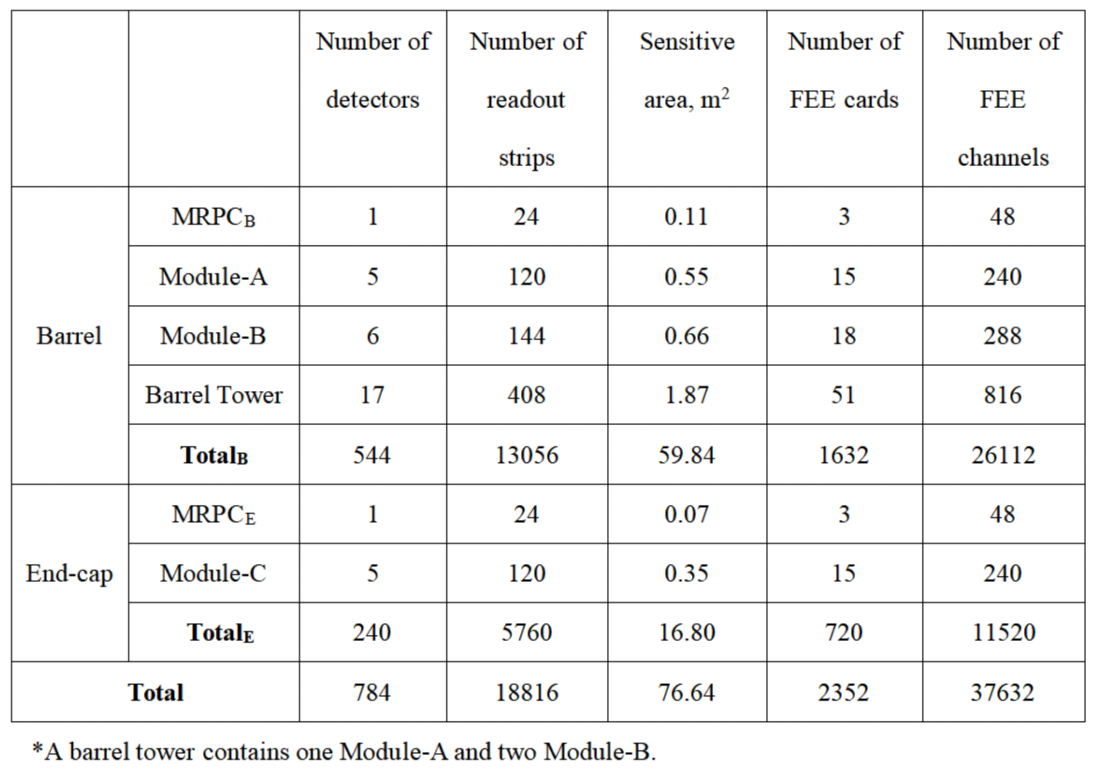}
    \caption{\label{TOF_Parameter}Main parameters of the MRPC-TOF system.}
\end{figure}

\begin{figure}[htbp]
    \centering
    \includegraphics[width=.7\textwidth]{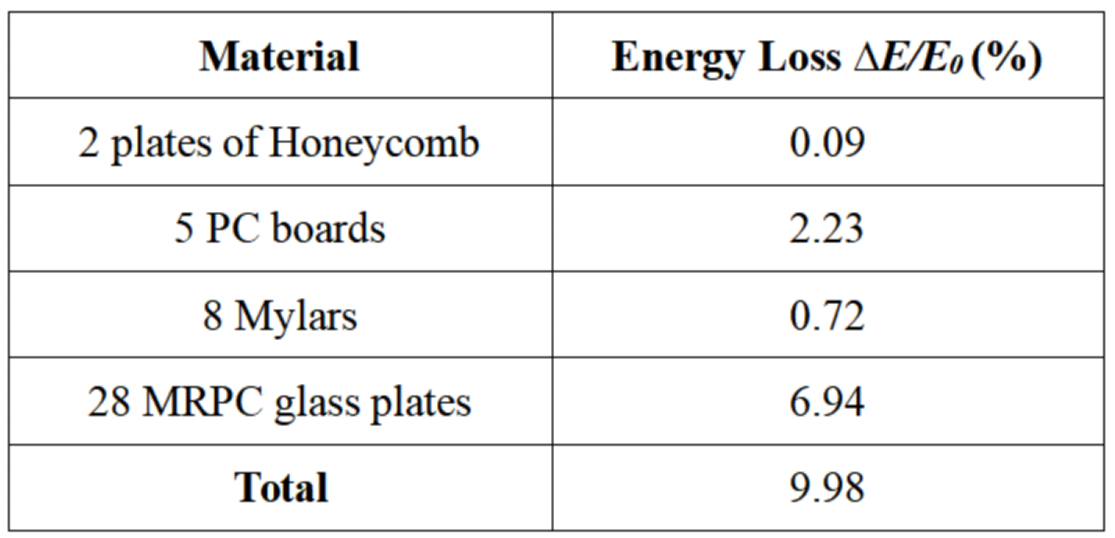}
    \caption{\label{TOF_X0}Material budget of the MRPC-TOF system.}
\end{figure}

Table~\ref{TOF_Parameter} summarizes the main parameters of the MRPC-TOF system. There are 4 chambers and 24 gas gaps in an MRPC. The thickness of the gas gap is 0.14\,mm and the thickness of the detector is 3.02\,cm. The total number of readout channels is estimated to be 37632, and the power consumption per channel is about 17\,mW. The material budget of the whole TOF system can be controlled below 0.1 radiation length (X$_0$), with detailed proportions shown in table~\ref{TOF_X0}. The total cost is estimated to be 5 million USD, with approximately 1 million USD spent on MRPCs. Such a 24 gas gaps MRPC-TOF with gas gap thickness of 0.14\,mm is anticipated to achieve an excellent time resolution $<$ 35\,ps . For the PID performance, a 3\,$\sigma$ separation of $\pi^{\pm}/K^{\pm}$ at 2.5\,GeV can be achieved.

\section{Conclusion}
\label{sec:Conclusion}

In the post-Higgs era, further scrutinizing the SM with precision much higher than the LHC is of great importance. Several future electron-positron Higgs factories, e.g. CEPC and FCC-ee, are proposed to achieve this goal. To fully accomplish scientific goals of these future colliders, the precise TOF measurement is critical. It can not only significantly improve the PID performance in the low momentum range, but also shows considerable potential in resolving ambiguities in the particle flow reconstruction. The MRPC, as an established TOF technology, has been widely applied in many fields. It features high detection efficiency, excellent time resolution, high rate capability, and comparatively low cost. We therefore propose a conceptual design of TOF system based on our independently developed sealed MRPC technology for future electron-positron Higgs factories. The MRPC-TOF is designed to be installed in front of the ECAL, covering an area of about 77\,m$^2$ in total. The total material budget of the whole TOF system can be controlled below 0.1\,X$_0$. The proposed design is expected to achieve a time resolution of less than 35\,ps, which can well meet the requirements of future Higgs factories for precise TOF measurement and efficient particle identification. The total cost is estimated to be 5 million USD, with approximately 1 million USD spent on MRPCs.

\section*{Acknowledgment}

The authors acknowledge the financial support provided in part by various funding agencies. This work is partially supported by the National Natural Science Foundation of China under grants No. 12042507, 11927901, 11420101004, 11461141011, 11275108, 11735009, and U1832118. The Ministry of Science and Technology of the People's Republic of China has also provided support under Grants No. 2020YFE0202001, 2018YFE0205200, and 2016YFA0400100. In addition, this work is supported by the International Partnership Program of the Chinese Academy of Sciences (Grant No. 113111KYSB20190030), the Innovative Scientific Program of the Institute of High Energy Physics.


\end{document}